\documentclass[twocolumn,showkeys,showpacs,preprintnumbers,prd,superscriptaddress,nofootinbib]{revtex4-1}
\usepackage{graphicx}
\usepackage{epsf}
\usepackage{bm}
\usepackage{amsmath}
\usepackage{amsfonts}
\usepackage{amssymb}
\usepackage{epstopdf}
\usepackage{color}
\usepackage{verbatim}
\usepackage{multirow}
\usepackage{bm}
\usepackage{soul}
\usepackage{hyperref}
\usepackage{booktabs}

\definecolor{darkpink}{rgb}{0.8, 0.0, 0.4}

\begin{document}

\title{
Possible detection of Baryon Acoustic Oscillations \texorpdfstring{\\}{ }
imprinted in the HI sources 
from the FASHI catalog
}

\author{Felipe Avila}
\email{felipeavila@on.br}
\affiliation{Observatório Nacional, Rua General José Cristino 77, 
São Cristóvão, 20921-400 Rio de Janeiro, RJ, Brazil}

\author{Camila Franco}
\email{camilafranco@on.br}
\affiliation{Observatório Nacional, Rua General José Cristino 77, 
São Cristóvão, 20921-400 Rio de Janeiro, RJ, Brazil}

\author{Armando Bernui}
\email{bernui@on.br}
\affiliation{Observatório Nacional, Rua General José Cristino 77, 
São Cristóvão, 20921-400 Rio de Janeiro, RJ, Brazil}

\begin{abstract}

Motivated by curiosity as to whether the observed Universe at low redshifts is well 
described by the concordance cosmological model, we investigate the large scale angular clustering of extragalactic neutral hydrogen HI sources in the local Universe. Using the second data
release of the FAST All Sky HI Survey (FASHI), we analyze a sample of 142,204 HI sources up to $z<0.09$, with a median redshift of $\bar{z} = 0.034$, covering $\sim 40\%$ of the sky. 
We compute the 2-point angular correlation
function, $w(\theta)$, and estimate its covariance using an analytical Gaussian model. 
We report the possible detection of a broad 
angular Baryon Acoustic Oscillation (BAO) signal, peaking near $50^\circ$. 
A comparison between a 5th-order polynomial fit and a quadratic baseline yields $\Delta \chi^2 = 9.06$ for $\Delta \mathrm{dof} = 3$ (difference of 3 degrees of freedom), corresponding to a nominal $1.9\,\sigma$ significance for a non-monotonic feature in $w(\theta)$ relative to the monotonic quadratic baseline, although this result alone does not establish a BAO detection. 
Our analyses show that, due to the projection effect, the expected transverse BAO scale shifts from $64^\circ$ in an infinitesimally thin redshift bin to $52^\circ$ in a bin of thickness $\Delta z = 0.09$, like the data we study, a value close to where our detection peaks, i.e., $\sim 50^\circ$. 
Furthermore, comparisons with an ensemble of lognormal mocks show that the observed broad bump is consistent with statistical fluctuations at the $1\sigma$ level, meaning that current uncertainties prevent us from unequivocally establishing it as a cosmological signal. 
Comprehensively, our study establishes a valuable observational baseline for wide-field HI surveys, showing the potential to trace primordial acoustic features even in the low redshift regime, where the transverse BAO signal is dominated by projection effects.

\end{abstract}




\maketitle


\section{Introduction}
\label{sec:introduction}

Over the past two decades, Baryon Acoustic Oscillations (BAO) have emerged as a primary science driver for major cosmological surveys~\citep{York2000, DES2016, DESI2013}. 
The reason for these initiatives is because 
the acoustic peak, observed in the 2-point correlation function at a characteristic scale of $\sim \! 100 \, h^{-1} \text{Mpc}$, 
is 
a robust standard ruler for testing cosmological models. 
Because the observed BAO scale depends on the Hubble expansion rate, $H(z)$, along the line of sight, and on the transverse comoving distance, $D_M(z)$, across the sky, it is highly sensitive to the expansion history of the Universe. 
Consequently, BAO measurements are valuable because they provide stringent constraints on the behavior of dark energy~\cite{Blake03,Seo03}. 

When combined with other cosmological studies and assuming a specific parameterization for the equation of state, recent results from the Dark Energy Spectroscopic Instrument (DESI) survey provide strong evidence for the evolution of dark energy equation of state, commonly referred to as dynamical dark energy~\cite{DESI:2025wyn,DESI:2025zgx}. 
However, the DESI measurements yield an isotropic BAO scaling parameter, $\alpha$, that is smaller than unity when compared to standard model predictions. Since this anomaly indicates shorter comoving distances at low redshifts, various alternative classes of models and parameterizations have been proposed to simultaneously explain the DESI observations and their combinations with other cosmological observables~\cite{Cortes:2024lgw,Giare:2024smz,DESI:2024aqx,DESI:2025fii,Miguel_ON}.

Motivated by these emerging challenges to the $\Lambda\text{CDM}$ paradigm and the recent DESI findings, it is important to investigate the BAO signal across current and upcoming generations of cosmological surveys. 
These efforts include analyses of new cosmic tracers, such as extragalactic neutral hydrogen (HI) sources~\cite{HIsurvey}.
To address this need, in this work we investigate large scale clustering within the FAST All Sky HI Survey (FASHI)~\cite{fashi_DR1,fashi_DR2}, a catalog of HI sources in the local Universe ($z < 0.09$) spanning approximately 
$19,500\,\text{deg}^2$. 
Recently, the second release of FASHI data enabled enhanced analyses, as it contains a larger number of HI sources corresponding to a more extensive observed area. 
With these data one can study largest angular separations between pairs in the 2-point angular correlation function, looking for the first time for robust evidences of the BAO signature in this cosmic tracer in the extreme local Universe. 

Several ongoing and planned radio facilities are dedicated to exploiting the 21 cm line as a cosmological probe. These include The Baryon Acoustic Oscillations from Integrated Neutral Gas Observations (BINGO)~\cite{bingo1,bingo2,bingo3}, 
and the Square Kilometre Array (SKA)~\cite{ska1,ska2,ska3} together with its pathfinder MeerKAT~\cite{meerkat1,meerkat2}, which will conduct large scale HI surveys. In parallel, intensity mapping experiments such as The Canadian Hydrogen Intensity Mapping Experiment (CHIME)~\cite{chime1,chime2,chime3}, The Hydrogen Intensity and Real-time Analysis eXperiment (HIRAX)~\cite{hirax}, and Tianlai~\cite{tianlai1,tianlai2} aim to map the large scale structure over broad redshift ranges, providing complementary constraints on dark energy and the expansion history of the Universe.

This work is organized as follows. 
Section~\ref{sec:data} introduces the FASHI data catalog. 
Section~\ref{sec:methodology} details the theoretical and computational framework employed to calculate the 2-point correlation function. 
Section~\ref{sec:Results} presents the main results, while Section~\ref{sec:final} is dedicated to a discussion of these findings alongside our final remarks.


\section{The FASHI catalog}
\label{sec:data}

In this work, we investigate observational data from the second data release\footnote{\url{https://zcp521.github.io/fashi.html}} of the FAST All Sky HI Survey (FASHI)~\cite{fashi_DR1,fashi_DR2}. 
Conducted with the Five-hundred-meter Aperture Spherical radio Telescope, FASHI provides a comprehensive census of 21-cm emission in the local Universe. The catalog comprises over 156,000 extragalactic HI sources distributed across a celestial footprint of approximately $19,500\,\text{deg}^2$, extending out to a redshift of $z = 0.09$~\footnote{The catalog provides velocities in the heliocentric reference frame. For the purposes of this work, the choice of reference frame does not affect either the results or their interpretation.}. By achieving a median sensitivity of 0.57 mJy beam$^{-1}$, this dataset probes a significantly larger comoving volume and reaches lower neutral hydrogen masses compared to legacy campaigns such as ALFALFA~\cite{alfalfa18}. Consequently, the FASHI catalog yields a highly complete and statistically robust sample of gas-rich galaxies, making it an optimal dataset for investigating the large scale structure and clustering properties of cold gas in the low redshift Universe~\cite{Wu24_isotropy}.

\begin{figure*}
\centering
\includegraphics[scale=0.6]{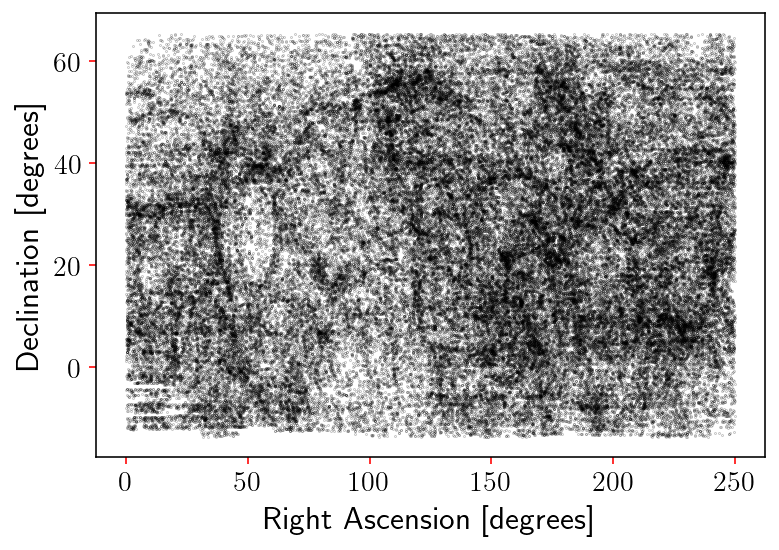}
\includegraphics[scale=0.6]{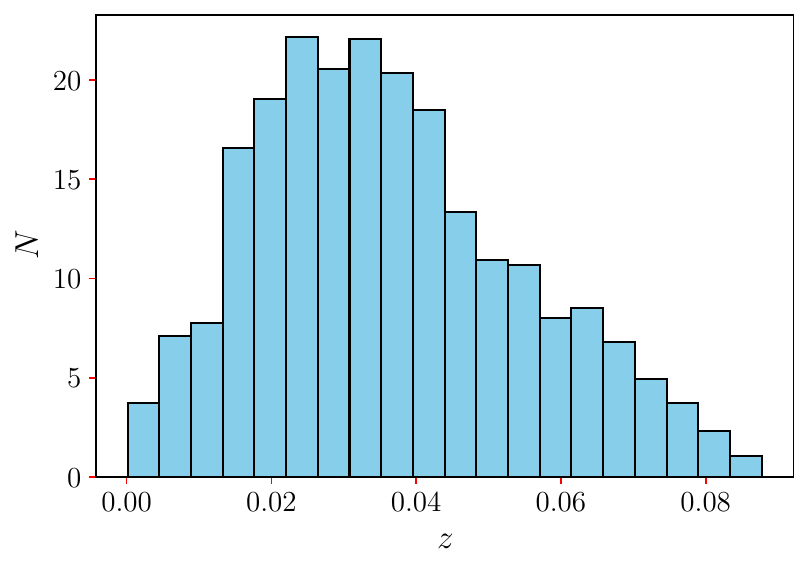}
\caption{\textit{Left panel:} The celestial footprint of the selected FASHI DR2 subsample, restricted to a contiguous region with right ascensions $0^\circ < \alpha < 250^\circ$ and declinations $-13.7^\circ < \delta < 65.4^\circ$. This area corresponds to a sky coverage fraction of $f_{\text{sky}} \simeq 0.4$. \textit{Right panel:} The normalized redshift distribution of the 142,204 extragalactic $\text{H\,\textsc{i}}$ sources ($z > 0$) retained for the clustering analysis, 
with median redshift $\bar{z} = 0.034$.}
\label{fig:footprint_histogram}
\end{figure*}

To maximize the statistical potential of the FASHI catalog, our primary analysis imposes only two conservative selection criteria: (i) a spatial footprint restricted to right ascensions $0^\circ < \alpha < 250^\circ$ and declinations $-13.7^\circ < \delta < 65.4^\circ$, and (ii) a cosmological cut requiring $z > 0$. This coordinate cut isolates a simple rectangular footprint, making it straightforward to generate the random catalog. The redshift condition excludes sources with negative radial velocities, effectively removing contamination from local high-velocity clouds (HVCs) and galactic objects. 
Figure~\ref{fig:footprint_histogram} summarizes the properties of this sample. 
The left panel illustrates the resulting footprint, which spans approximately $40\%$ of the celestial sphere ($f_{\text{sky}} \simeq 0.4$), while the right panel presents the redshift distribution of the final 142,204 selected HI sources, characterized by a median redshift of $\bar{z} = 0.034$. 
Given these settings, the comoving volume of the sample is $V \approx 0.03\;h^{-3}\,\text{Gpc}^3$. 
This is about a factor of 30 smaller than the volume typically required for a robust 3D BAO detection, i.e., $V \sim 1\;h^{-3}\,\text{Gpc}^3$~\cite{Miller:2001cf,SDSS:2005xqv}. 
Consequently, a 3D BAO measurement is not feasible with this sample. 
Our analysis is therefore restricted to search for the transverse BAO signal. 
Additional selection criteria will be introduced later in the text as required by the analysis.

\section{The 2-point angular correlation function}
\label{sec:methodology}

In this section, we will present the theoretical model and the estimator for The 2-point angular correlation function (2PACF).

\subsection{Theoretical model}
\label{sec:model}

The observed angular galaxy density fluctuation along a direction $\hat{\mathbf{n}}$ is the line-of-sight projection of the spatial density field $\delta_g(\mathbf{x}, z)$~\cite{Peebles80,Crocce2011} 
\begin{equation}
    \delta_g(\hat{\mathbf{n}}) = \int dz \, \phi(z) \delta_g(\mathbf{x}, z),
\end{equation}
where $\phi(z)$ is the survey's radial selection function. 

Decomposing this field into spherical harmonics, its variance as a function of scale is characterized by the angular power spectrum, $C_\ell$. In linear perturbation theory, $C_\ell$ is related to the dimensionless primordial curvature power spectrum, $P_\Phi(k)$, through the integration over wavenumbers $k$~\cite{Chisari19} 
\begin{equation}
    C_\ell = \frac{2}{\pi} \int dk \, k^2 P_\Phi(k) |\Delta_\ell(k)|^2.
\end{equation}
The total transfer function, $\Delta_\ell(k)$, encompasses both the intrinsic matter density fluctuations and the Redshift-Space Distortions (RSD). Assuming a linear galaxy bias $b(z)$, these contributions can be compactly expressed as~\cite{Chisari19} 
\begin{equation}
\begin{split}
    \Delta_\ell(k) &= \int dz \, \phi(z) \bigg[ b(z) T_\delta(k,z) j_\ell(kr) \\
    &\quad + \frac{1+z}{H(z)} T_\theta(k,z) j''_\ell(kr) \bigg],
\end{split}
\end{equation}
where $H(z)$ is the Hubble parameter, $r$ is the comoving distance, $T_\delta$ and $T_\theta$ are the transfer functions for the matter overdensity and the velocity field divergence, respectively, and $j_\ell$ is the spherical Bessel function of order $\ell$. 
The term proportional to the second derivative with respect to its argument $kr$, $j''_\ell(kr)$, modulated by the kinematic factor $(1+z)/H(z)$, accounts for the RSD induced by coherent galaxy infall.

Finally, the 2PACF is obtained by transforming the angular power spectrum to configuration space via the Addition Theorem~\cite{Peebles80}:
\begin{equation}
    w(\theta) = \sum_{\ell} \frac{2\ell+1}{4\pi} C_\ell P_\ell(\cos\theta),
\end{equation}
where $P_\ell(\cos\theta)$ are the Legendre polynomials of degree $\ell$, and $\theta$ is the angular separation between two lines of sight. 

To perform these numerical integrations without relying on the Limber approximation, we utilize the Core Cosmology Library (\textsc{CCL})\footnote{\url{https://github.com/LSSTDESC/CCL}} software package~\cite{Chisari19}. We adopt the Planck 2018 cosmological parameters ($\Omega_c = 0.2656$, $\Omega_b = 0.0494$, $h = 0.6727$, $\sigma_8 = 0.8120$, $n_s = 0.9649$)~\cite{Planck20}, which yield a sound horizon scale of $r_d = 147.05 \, \text{Mpc}$. We compute the non-linear matter power spectrum using the halofit prescription~\cite{Takahashi:2012em}. Although halofit is not optimized for modeling BAO damping, non-linear gravitational effects are subdominant compared to the projection effects introduced by the redshift shell. Consequently, employing a more complex non-linear model would not alter the interpretation of our results.

\subsection{Landy-Szalay estimator}
\label{sec:LS_estimator}

To quantify the spatial distribution and clustering properties of sources across the celestial sphere, we compute the 2-point angular correlation function (2PACF), denoted as $\omega(\theta)$. To minimize edge effects and properly account for the survey's complex mask, footprint, and varying sensitivity, we employ the robust Landy-Szalay estimator, defined as~\cite{Landy93}:
\begin{equation}
    \omega(\theta) = \frac{DD(\theta) - 2DR(\theta) + RR(\theta)}{RR(\theta)},
\end{equation}
where $DD(\theta)$, $DR(\theta)$, and $RR(\theta)$ represent the normalized pair counts of data-data, data-random, and random-random objects, respectively, falling within an angular separation bin $\theta$. The random catalog was constructed to faithfully match the angular selection function and footprint geometry. It has twenty more points than the data to reduce the shot-noise\footnote{As seen in Figure~\ref{fig:footprint_histogram}, the angular distribution is non-uniform due to regions observed multiple times~\cite{fashi_DR2}; therefore, the random catalog must reproduce this pattern to avoid overestimating the 2PACF amplitude.}. In appendix \ref{app:appendix_null_test}, we verified if this random catalog biased our analysis.

The angular separation $\theta$ between any two objects located at coordinates given by right ascensions ($\alpha_1, \alpha_2$) and declinations ($\delta_1, \delta_2$) is calculated via the spherical law of cosines 
\begin{equation}
\theta = \arccos \left[ \sin\delta_1 \sin\delta_2 + \cos\delta_1 \cos\delta_2 \cos(\alpha_1 - \alpha_2) \right]. 
\end{equation}
To calculate the 2PACF from the data, we utilize the open-source \texttt{TreeCorr}\footnote{\url{https://rmjarvis.github.io/TreeCorr}}~\cite{Jarvis04}. Because our analysis spans exceptionally broad angular scales, the flat-sky approximation is invalid. We therefore configure \texttt{TreeCorr} to use the \texttt{Arc} metric rather than its default Euclidean setting, ensuring that all pair separations are rigorously computed using exact great circle distances.

\subsection{Covariance matrix}
\label{sec:cov_matrix}

To estimate the statistical uncertainties of our 2PACF measurements, we employ an analytical Gaussian covariance matrix~\cite{Crocce2011}. While generating mock catalogs is a standard procedure for covariance estimation, constructing realistic HI mocks at very low redshift ($z < 0.09$) is notoriously difficult due to the complex, non-linear local structures inherent to the local Universe. The analytical Gaussian matrix has been extensively tested and validated against $N$-body simulations in previous literature~\cite{Crocce2011,DES:2024pwq,DES:2026aht}. We acknowledge, however, that this theoretical baseline may underestimate the true error.

The theoretical Gaussian covariance matrix accounts for partial sky coverage ($f_{\text{sky}}$) and couples the shot noise term ($1/\bar{n}$) with the fiducial clustering signal. It is defined as~\cite{Crocce2011}:
\begin{equation}\label{eq:cov_matrix}
\begin{split}
\text{Cov}(\theta_i, \theta_j) &\equiv \frac{2}{f_{\text{sky}}} \sum_{\ell} \frac{2\ell+1}{(4\pi)^2} \\
&\quad \times P_\ell(\cos\theta_i) P_\ell(\cos\theta_j) \left( C_\ell + \frac{1}{\bar{n}} \right)^2,
\end{split}
\end{equation}
where the $P_\ell(\cos\theta)$ are evaluated at the centers of the respective angular bins. 
We evaluate the power spectrum $C_\ell$ using the Planck 2018 cosmology described in the previous section. 
For this calculation, we adopt the median redshift of our sample, i.e. $\bar{z} = 0.034$, and assume a linear bias $b = 1$~\cite{Martin12,Papastergis:2013lia}. 
Based on the spatial footprint and the size of the FASHI DR2 sample, we set the sky fraction to $f_{\text{sky}} = 0.4$ and the surface number density to $\Sigma = 28,437 \, \text{sr}^{-1}$. 
The theoretical error will also depend on the distribution of HI sources, $N(z)$. 

\begin{figure}
    \centering
    \includegraphics[scale=0.6]{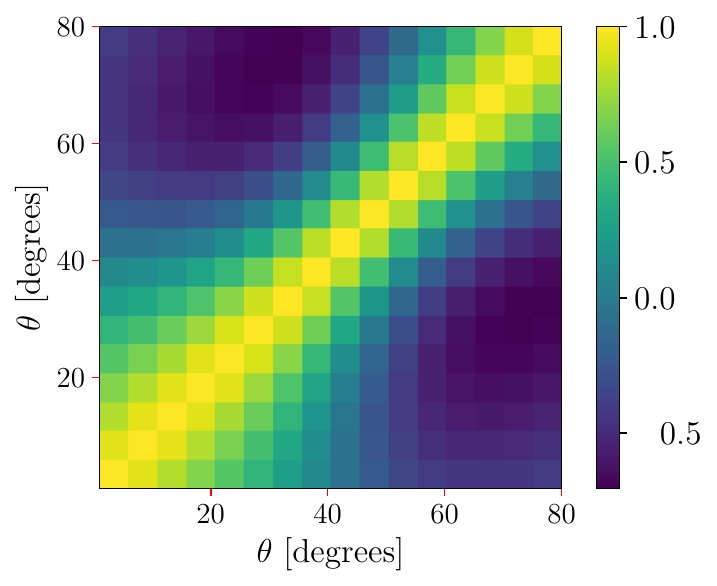}
    \caption{Reduced theoretical correlation matrix for the 2PACF. The matrix is derived from the analytical Gaussian covariance model, coupling cosmic variance and shot noise for the FASHI DR2 sample parameters. The color bar indicates the correlation coefficient between different angular separation bins.}
    \label{fig:cov_matrix}
\end{figure}

Figure~\ref{fig:cov_matrix} displays the corresponding correlation matrix derived from this theoretical covariance. The specific angular scales $\theta$ chosen to evaluate the 2PACF will be presented and discussed in the next section.

\section{Results}
\label{sec:Results}

The maximum angular scale for evaluating $w(\theta)$ is determined by the expected BAO scale in a given fiducial model. In a standard 3D analysis, for instance, capturing a BAO scale of $100 \, h^{-1}\text{Mpc}$ typically requires computing the correlation function up to $150$ or $200 \, h^{-1}\text{Mpc}$~\cite{Beutler11,DESI25}, depending on the survey volume. Similarly, the chosen bin width dictates the noise level of the extracted BAO signal~\cite{deCarvalho18}.

In a purely angular analysis, the transverse BAO scale is defined as 
\begin{equation}\label{eq:bao_scale}
\theta_{\rm BAO}(z) \equiv \frac{r_d}{(1+z)D_{A}(z)} \,,
\end{equation}
where $D_A(z)$ is the angular diameter distance. In the low redshift regime, $D_A(z) \approx cz/H_0$. 
At $z = 0.034$, assuming $H_0 = 67.27~\text{km s}^{-1}\text{Mpc}^{-1}$, we obtain $D_A = 151.90 \, \text{Mpc}$. 
Using the Planck sound horizon of $r_d = 147.05 \, \text{Mpc}$, this yields $\theta_{\rm BAO}(z=0.034) = 53.54^\circ$. 
Since increasing $H_0$ decreases $D_A$, and consequently increases $\theta_{\rm BAO}$, our evaluation must extend beyond $53^\circ$. 
Given that our sample is limited by a declination span of $80^\circ$, we adopt this as the maximum angular scale. 
Appendix \ref{app:density_pairs} confirms that the pair density in the interval $[0^{\circ}, 80^{\circ}]$ is sufficient to produce a detectable signal. Furthermore, strong projection effects~\cite{Simpson09,Sanchez11} at low redshifts shift the BAO feature to smaller angles, ensuring the expected peak remains below $53^\circ$. Unless otherwise specified, we compute $w(\theta)$ between $1^\circ$ and $80^\circ$ using linear $\Delta\theta=5^\circ$ bins.

\begin{figure*}
\centering
\includegraphics[scale=0.6]{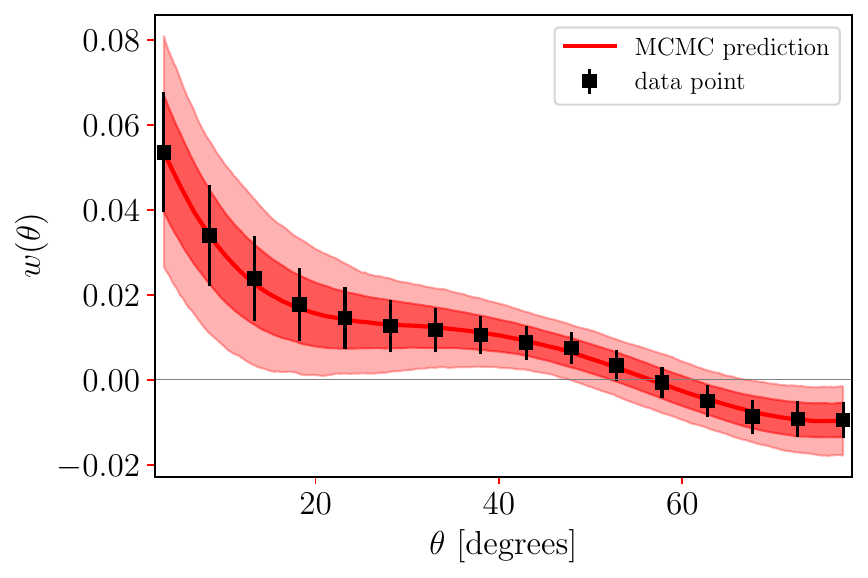}
\includegraphics[scale=0.6]{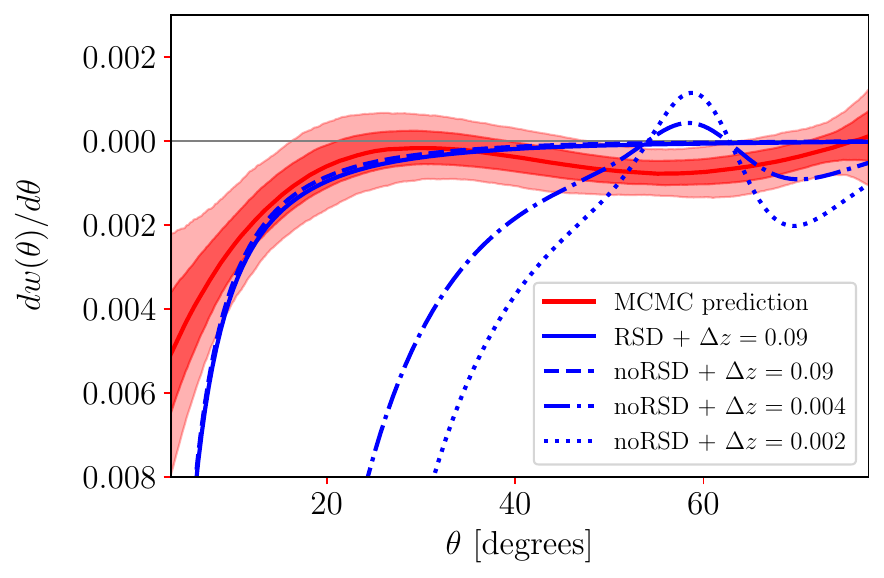}
\caption{\textit{Left panel:} The 2PACF, $w(\theta)$, of the FASHI sample at a median redshift of $\bar{z} = 0.034$. Black squares represent the data points measured via the Landy-Szalay estimator, with error bars given by the square root of the diagonal elements of the theoretical covariance matrix. The solid red line and shaded regions indicate the best-fit 5th-order polynomial and its associated confidence intervals. A broad bump is distinctly visible between $20^\circ$ and $60^\circ$. 
\textit{Right panel:} The first derivative of the 2PACF, $dw/d\theta$, as a function of the angular scale $\theta$. The solid red line and shaded regions represent the mean prediction and confidence intervals ($1\sigma$ and $2\sigma$) from the MCMC analysis. The blue lines show the theoretical expectations for four scenarios: RSD $+$ $\Delta z = 0.09$ (solid line), no RSD $+$ $\Delta z = 0.09$ (dashed line), no RSD $+$ $\Delta z = 0.004$ (dash-dotted line), and no RSD $+$ $\Delta z = 0.002$ (dotted line). While the thin-shell model without RSD ($\Delta z = 0.002$) predicts a clear BAO signature with a zero crossing near $60^\circ$, projection effects in the finite-shell models suppress this feature.
}
\label{fig:2PACF}
\end{figure*}

In Figure~\ref{fig:2PACF} (left), we present the 2PACF of the FASHI sample computed using the Landy-Szalay estimator. A broad bump is clearly visible between $20^\circ$ and $60^\circ$. The error bars are derived from the diagonal elements of the theoretical covariance matrix (see equation~\ref{eq:cov_matrix}). 
Additionally, we model the data using a 5th-order polynomial fit, as detailed below.

To extract cosmological information from the angular BAO bump, two approaches are typically adopted: (i) fitting the bump using a dilation parameter, $\alpha$, based on a fiducial model (as described in Section \ref{sec:model})~\cite{Abbott24, DESI25}; 
or (ii) extracting the information directly via an empirical parameterization~\cite{Sanchez11,deCarvalho21}. 
This second method is particularly effective for thin redshift shells, as it eliminates the need to correct for projection effects~\cite{Menote22,Avila:2025qxc}. 
In this study, however, we opt for an illustrative analysis of the BAO feature by modeling the data with a 5th-order polynomial fit. 
It has been shown that a polynomial of this order fits the correlation function well~\cite{Anselmi:2017cuq,Anselmi:2017zss}. 
The projection effect in our analyses is truly significant, as shall discuss in section~\ref{sec:proj_effect} where we compare the expected angular positions of the 
BAO signal in an infinitesimally thin and in a thick redshift bins. 

To investigate the variation of the correlation function with angular scale, we compute its first derivative, $dw/d\theta$. The 5th-order polynomial is fitted to the $w(\theta)$ data using a Markov Chain Monte Carlo (MCMC) approach implemented via the \texttt{emcee} package\footnote{\url{https://emcee.readthedocs.io/en/stable/}} \cite{Foreman-Mackey_2013} to properly sample the parameter space and estimate uncertainties. The full posterior distributions of the fitted polynomial coefficients are detailed in Appendix \ref{sec:appendix_mcmc}. Theoretically, this derivative is expected to cross zero twice (for an infinitesimally thin shell), exhibiting a profile similar to a normal curve intersecting the zero axis, where the second crossing corresponds to the BAO scale. As shown in the Figure~\ref{fig:2PACF} (right), the derivative of our polynomial fit approaches zero between $20^\circ$ and $40^\circ$ within a $2\sigma$ confidence interval.

In the Figure~\ref{fig:2PACF} (right), we plot the theoretical expectation for $dw/d\theta$ under three distinct scenarios: (i) a thin redshift shell without RSD, (ii) a finite shell with RSD, and (iii) a finite shell without RSD. In the first case, represented by the blue dotted and dash-dotted line, clear zero-crossings are visible, with the BAO scale emerging at the second crossing near $60^\circ$. 
However, in the latter two theoretical scenarios, the severe projection effect completely suppresses the BAO signature. 
Consequently, the theoretical model indicates that, given our sample's median redshift and shell width, a BAO bump should not be observable. 
The theoretical curve asymptotically approaches zero. Notably, between $50^\circ$ and $60^\circ$, the polynomial derivative exhibits a marginal dip below the zero line. Though the statistical significance of this feature is low, it qualitatively traces the expected behavior of a BAO signal, where the correlation function drops and produces a negative gradient similar to the theoretical thin-shell scenario.

\begin{figure}
\centering
\includegraphics[scale=0.6]{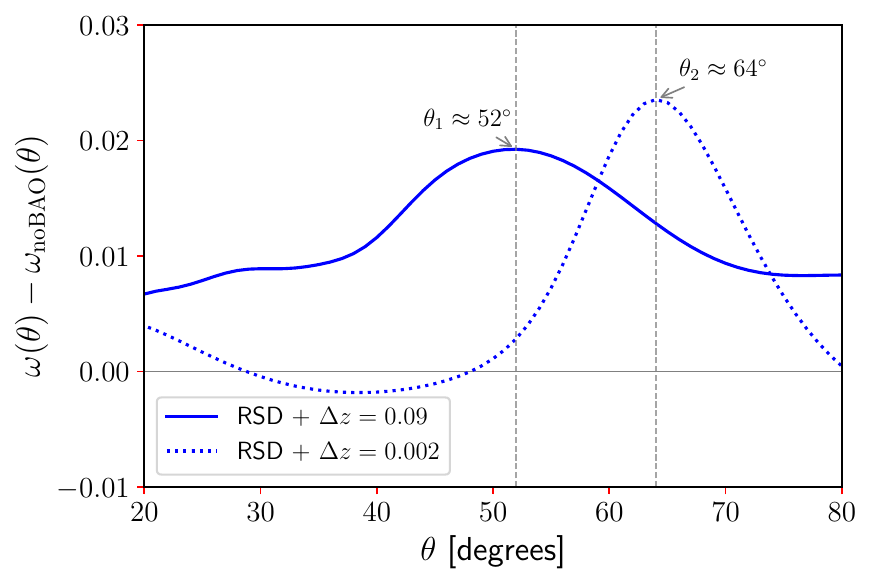}
\caption{
Theoretical BAO residual $w(\theta) - w_{\rm noBAO}(\theta)$, for two configurations. 
The solid line shows the realistic configuration of our analysis (RSD $+$ $\Delta z = 0.09$), peaking at $\theta_1 \simeq 52^\circ$; its amplitude has been multiplied by a factor of $200$ for visualization purposes only, so that the two curves can be displayed on the same vertical scale. 
The dotted line shows the idealized thin-shell case (RSD $+$ $\Delta z = 0.002$), peaking at $\theta_2 \simeq 64^\circ$. Therefore, the theoretical projection-induced angular shift is $\Delta\theta \simeq 12^\circ$, corresponding to a $19\%$ angular shift of the BAO signal.}
\label{fig:proj_theory}
\end{figure}

To quantify the impact of line-of-sight projection on the transverse BAO feature, 
we compute the BAO residual, defined as the difference between the 2PACF 
computed with and without BAO wiggles in the transfer function~\cite{Einsenstein98}, that is, we calculate: 
$w(\theta) - w_{\rm noBAO}(\theta)$. 
Figure~\ref{fig:proj_theory} shows this BAO residual for two configurations: 
an idealized thin shell ($\Delta z = 0.002$) 
with RSD effect, and the realistic 
FASHI configuration including RSD and the 
thick-shell case corresponding to our data analysis ($\Delta z = 0.09$). 
In the thin-shell case, the residual exhibits a clear peak at $\theta_2 \simeq 64^\circ$, which is the transverse BAO scale expected by the $\Lambda$CDM model in this idealized scenario. 
In the thick-shell case, the projection effect along the line of sight displaces and broadens the BAO feature~\citep{Crocce08}, shifting the peak to $\theta_1 \simeq 52^\circ$. 
The expected angular shift, due to the projection effect, is therefore $\Delta\theta \approx 12^\circ$, corresponding to a $\sim 19\%$ shift of the BAO signal. 
In other words, in the analyses of FASHI data, 
in a thick redshift shell with $\Delta z = 0.09$ 
and $\bar{z} = 0.034$, one should expect the detection of the transverse BAO signal at $\sim 52^\circ$.

\subsection{Statistical significance of the BAO signal}
\label{sec:significance}

The inspection of the 2PACF $w(\theta)$ and its derivative, shown in the plots of Figure~\ref{fig:2PACF}, suggests the presence of a broad, non-monotonic feature between $20^\circ$ and $60^\circ$. 
To quantify whether this structure is statistically significant, we first investigate the standard empirical template of~\cite{Sanchez11}, consisting of a power-law baseline supplemented by a Gaussian bump. 
This model, however, proved inadequate for our dataset. 
The BAO feature we observe is much broader than the narrow Gaussian typically used to describe the BAO signal, and its shape overlaps significantly with the power-law baseline. 
In consequence, the MCMC posteriors for both BAO features, i.e., the bump position and its width, remain essentially flat, and the Gaussian term does not improve the fit of $w(\theta)$ 
in a meaningful way.

We therefore adopt a different strategy to compute the statistical significance of the BAO signal in the HI sample from FASHI. 
This time we fit the data with two nested models: a 5th-order polynomial, which has been shown to reproduce $w(\theta)$ of large-scale structure samples accurately~\cite{Anselmi:2017cuq, Anselmi:2017zss} and is flexible enough to capture the broad BAO bump, 
and a quadratic polynomial which represents a strictly smooth baseline with no BAO feature. 
Because the quadratic polynomial is a special case of the 5th-order polynomial, obtained by setting the three highest-order coefficients to zero, the improvement in $\chi^2$ directly measures how much the additional freedom is required by the data. 
Both models are sampled with the same \texttt{emcee} setup described above, using the full covariance matrix of the data. 
The best-fit parameters and posterior distributions are reported in Appendix~\ref{sec:appendix_mcmc}.

Figure~\ref{fig:fit_poly25} shows our result. 
The top panel displays the FASHI 2PACF together with the best-fit 5th-order polynomial (solid red curve) and the quadratic baseline (dashed blue curve). 
The bottom panel shows the residuals with respect to the quadratic baseline, with error bars that propagate both the
measurement uncertainty and the uncertainty on the quadratic fit itself. 
The 5th-order polynomial systematically deviates from the quadratic baseline in the angular interval $40^\circ$ -- $60^\circ$, where the broad bump appears, while in the rest of the 
angular range investigated the two models agree within $1\,\sigma$ uncertainty.

Interestingly, the difference of these two curves reaches its maximum near $50^\circ$, very close to where the thick-shell model predicts the projection-displaced 
BAO feature to appear ($\sim 52^\circ$, see Figure~\ref{fig:proj_theory}). 
This agreement indicates that the broad structure seen in the data may indeed be a shifted BAO signature, rather than a random fluctuation of the smooth baseline. 
However, we emphasize that this is only a qualitative comparison, since the residual alone does not allow us to measure the BAO scale.

To quantify the performance of these fits, we compare both and find an improvement of 
$\Delta \chi^2 = 9$ when moving from the quadratic to the 5th-order polynomial. 
This corresponds to a nominal significance of around $1.9\,\sigma$. 
We emphasize that this value quantifies the presence of some signature in $w(\theta)$ 
not well described by a quadratic polynomial fit, 
and does not necessarily imply the detection of a BAO signal.

 \begin{figure}
\centering
\includegraphics[scale=0.42]{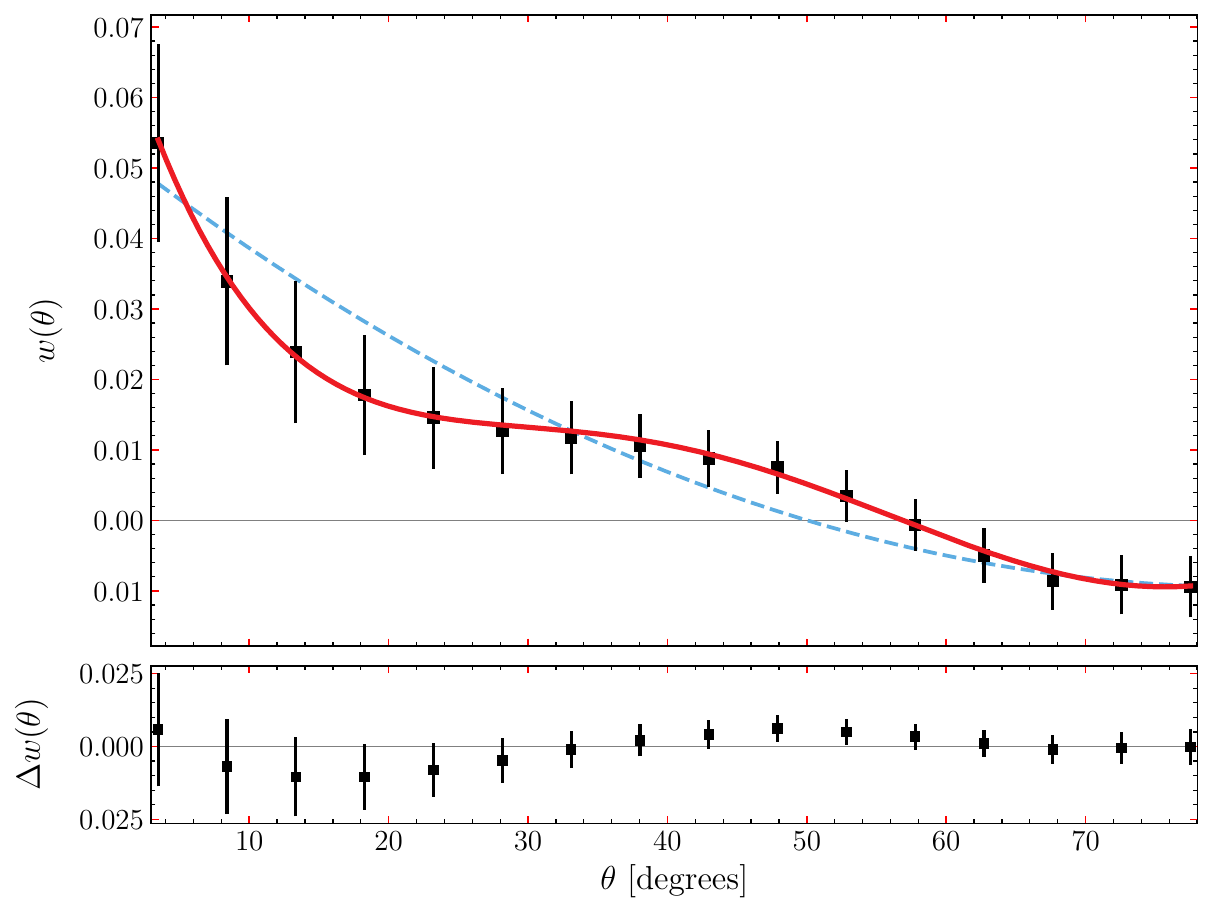}
\caption{\textit{Top panel:} 
The 2PACF from our analyses of the FASHI data, 
that is, the black squares data points and their error bars are the same as those shown in the left panel of Figure~\ref{fig:2PACF}, and this time our study compares two best-fit procedures using a 5th-order polynomial versus a second-order polynomial.
The solid red curve is the best-fit 5th-order polynomial, and the dashed blue curve is the best-fit quadratic, used as a smooth baseline. 
\textit{Bottom panel:} Residuals with respect to the quadratic baseline. Error bars include both the measurement uncertainty and the propagated uncertainty on the quadratic fit. 
The systematic deviation of the data from zero between $\theta \simeq 40^\circ$ and $60^\circ$, with maximum near $50^\circ$, traces the broad bump seen in the top panel and is consistent with the theoretical thick-shell prediction (see Figure~\ref{fig:proj_theory}).}
\label{fig:fit_poly25}
\end{figure}

\section{Robustness Tests, Discussions, and Final Remarks}
\label{sec:final}

Since the theoretical model demonstrates that the BAO signature is completely suppressed given our sample's median redshift and shell width ($\Delta z = 0.09$), one might suspect that a systematic error in the data is driving the anomalous amplitude of the 2PACF. However, there are no known systematic effects capable of smoothly and continuously increasing the correlation amplitude across such broad angular scales. Known physical structures, such as clusters and voids, typically manifest on much smaller angular scales~\cite{Avila18,Franco24}. Next, we will discuss some possible systematics.

\subsubsection{Zone of Avoidance}
\label{sec:zoa}

A common practice in cosmological analyses is to exclude observations located within or near the Zone of Avoidance (ZoA)~\cite{Novaes:2018kbw}. However, given the sensitivity of the FASHI survey~\cite{fashi_DR1,fashi_DR2} and the fact that Galactic dust does not attenuate the radio signal~\cite{Wilson13}, there is no justification for removing these sources from our catalog.

Nevertheless, to test the robustness of our signal against potential Galactic foregrounds, 
we exclude sources within the ZoA band. 
We analyze three masking thresholds based on latitude Galactic cuts: (i) $|b| < 5^\circ$, 
(ii) $|b| < 10^\circ$, and (iii) $|b| < 15^\circ$\footnote{For each threshold, both the theoretical covariance matrix and the random catalog must be recalculated. The covariance depends on the updated surface number density and sky fraction, while the random catalog must precisely reflect the newly masked footprint.}. Table~\ref{tab:galactic_cuts} summarizes the properties of the resulting masked catalogs, including the number of remaining sources ($N$), the updated fractional sky coverage ($f_{sky}$), and the surface number density ($\Sigma$). The random catalog is modified in the same way as the data. When calculating the theoretical error, we hold the median redshift and the bias.

\begin{table}[h]
\centering
\caption{Summary of the catalog properties after applying different Galactic latitude ($|b|$) masks. The columns list the masking threshold, the number of remaining galaxies ($N$), the fractional sky coverage ($f_{sky}$), and the surface number density of HI sources per sr ($\Sigma$).}
\label{tab:galactic_cuts}
\begin{tabular}{lccc}
\hline
\hline
\textbf{Cut} & $\mathbf{N}$ & $\mathbf{f_{sky}}$ & $\mathbf{\Sigma}$ \textbf{[sr$^{-1}$]} \\ 
\hline
Base sample (no $|b|$ cut) & 142,204 & 0.40 & 28,437 \\ 
$|b| > 5^\circ$            & 136,319 & 0.38 & 28,564 \\ 
$|b| > 10^\circ$           & 130,352 & 0.36 & 28,916 \\ 
$|b| > 15^\circ$           & 123,809 & 0.34 & 29,179 \\ 
\hline
\hline
\end{tabular}
\end{table}

\begin{figure}
\centering
\includegraphics[scale=0.6]{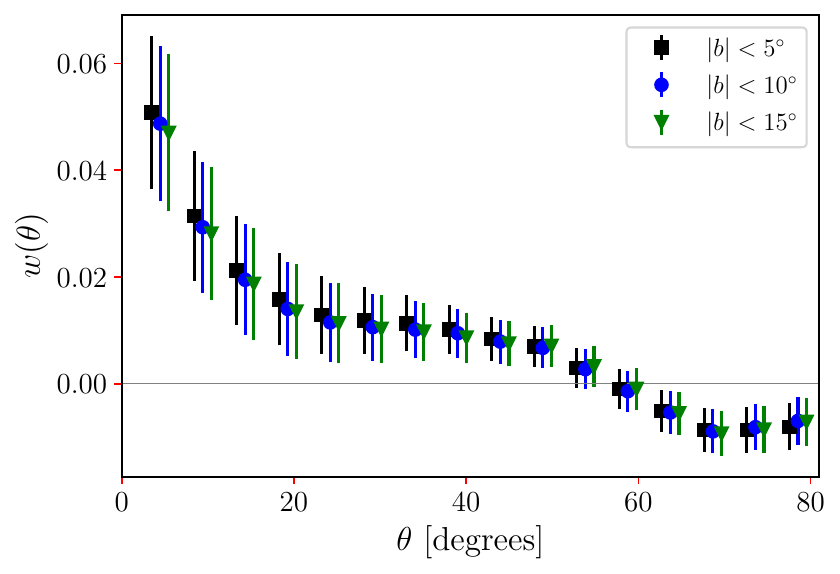}
\caption{The 2PACF evaluated under three different Zone of Avoidance (ZoA) masking thresholds: $|b| < 5^\circ$, $|b| < 10^\circ$, and $|b| < 15^\circ$. The persistence of the broad bump in all scenarios demonstrates that the signal is isotropic and independent of the Galactic plane. For visual clarity, the data points have been horizontally offset by $1^\circ$ relative to one another.}
\label{fig:mask}
\end{figure}

Figure~\ref{fig:mask} displays the 2PACF for these three cases. The results demonstrate that the broad bump is independent of the excluded Galactic region. This stability reinforces the hypothesis that the observed bump is of cosmological origin, as a genuine physical signal must be isotropic and insensitive to the specific survey footprint.

\subsubsection{Projection Effect}
\label{sec:proj_effect}

Another potential source of systematic error is the projection effect. Because our analysis employs a finite redshift shell rather than a thin slice, the line-of-sight projection of distant HI sources could potentially induce an artificial bump in the 2PACF.

\begin{figure*}
\centering
\includegraphics[scale=0.6]{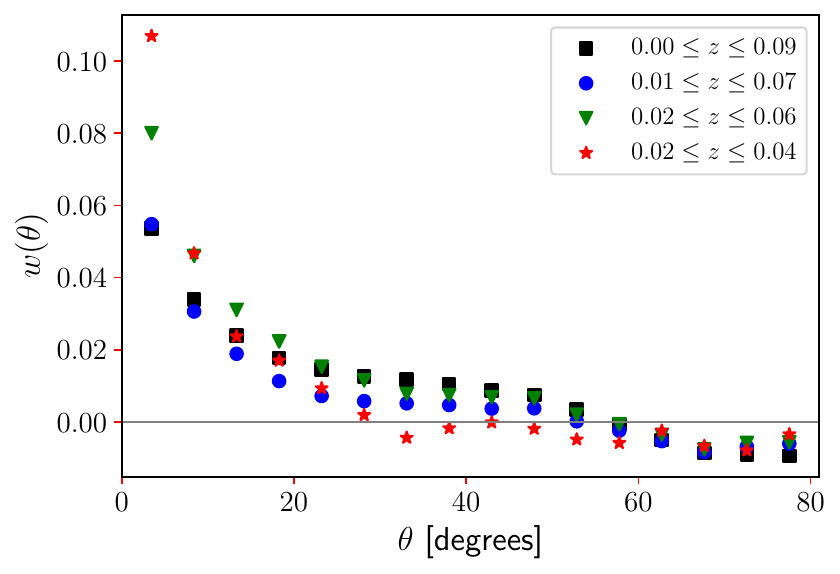}
\includegraphics[scale=0.6]{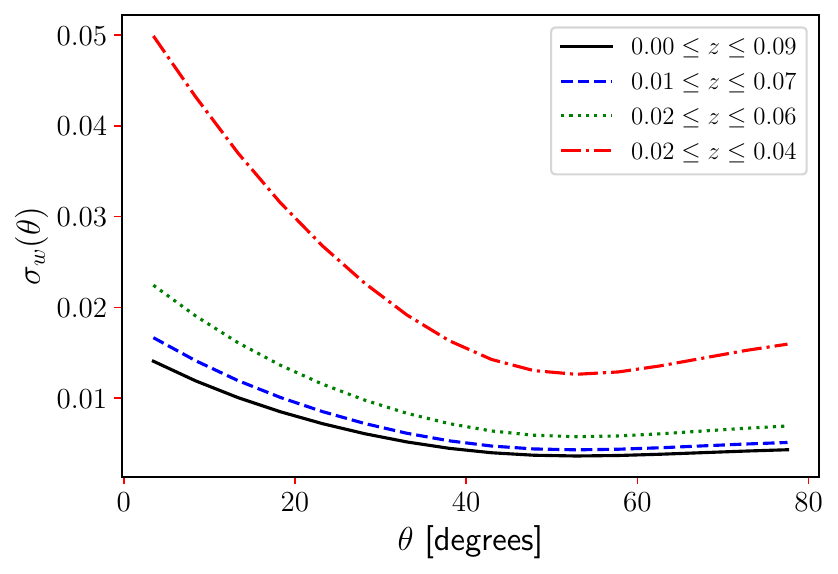}
\caption{\textit{Left panel:} The 2PACF evaluated across four redshift shells with different thicknesses. As the shell narrows, $w(\theta)$ drops more steeply at small angles, while in the bump region the four measurements remain consistent within the 
theoretical uncertainties. The predicted shift of the acoustic feature toward larger angular scales cannot be resolved with the current statistical power. \textit{Right panel:} Theoretical error for each bin. Note how the error increases substantially as we narrow the redshift interval, reducing the sensitivity to the 
projected BAO feature.}
\label{fig:bin_test}
\end{figure*}

To test this hypothesis, we systematically reduce the redshift shell thickness and calculate 
the 2PACF in several cases. 
Figure~\ref{fig:bin_test} (left) displays the results for four concentric redshift bins: 
$0.00 \leq z \leq 0.09$ (main sample), $0.01 \leq z \leq 0.07$, $0.02 \leq z \leq 0.06$, and $0.02 \leq z \leq 0.04$, which correspond to samples with decreasing number of HI sources, namely with a number of sources $10\%$, $33\%$, and $58\%$ lower, respectively, compared to the main sample. 
The most visible effect of narrowing the shell is a steeper drop of $w(\theta)$ at small angular scales, while in the region of the broad bump, i.e., $40^\circ$--$60^\circ$, the four measurements overlap within the theoretical uncertainties, as observed in the right panel of Figure~\ref{fig:bin_test}~\footnote{We fixed the bias and varied the surface density and the median redshift.}. 
In fact, as the redshift shell narrows, the theoretical errors increase substantially, 
reflecting the reduced statistical power of each subsample. In summary, this binning test illustrates what happens when one searches for the BAO signal in thick redshift bins.

We now examine in detail whether the observed broad BAO signature appears at the angular scale predicted by the BAO phenomenon considering a thick redshift shell data analysis. 
As shown in Figure~\ref{fig:proj_theory}, the thin-shell expectation places the BAO feature around $64^\circ$, while the thick-shell configuration that matches our sample 
(RSD $+$ $\Delta z = 0.09$) shifts it to $\sim 52^\circ$. 
The observed maximum of the broad signal, found around $50^\circ$, is therefore in qualitative agreement with the thick-shell prediction. 
Due to the projection effect,  this angular shift in our data analysis is of $12^\circ$, 
or $19\%$. 
For the sake of comparison, the authors of ref.~\cite{Sanchez11} found a $\sim 10\%$ shift for 
a shell of width $\Delta z = 0.1$ at $z = 0.3$, while the authors of ref.~\cite{Ribeiro25} 
reported $\sim 23\%$ for a shell of width $\Delta z = 0.07$ at $z_{\rm eff} = 0.075$. 
Our prediction of $\sim 19\%$ at $z = 0.034$ fits naturally into this picture: as the redshift decreases and the shell thickness grows, projection effects become more pronounced.

The binning test presented in Figure~\ref{fig:bin_test} is consistent with this 
interpretation, where, in fact, the most robust feature of the test is the change in 
the shape of $w(\theta)$ at small angular scales, which becomes steeper as the shell narrows, 
as expected if the smearing from the projection effect is being reduced. 
In the bump region, however, the four bins agree within the theoretical uncertainties, 
and the predicted shift toward larger angles cannot be resolved with the current statistical power. 
This coherence between the data and the thick-shell scenario is encouraging, but it also highlights an important limitation: the projection effect is large enough that analytical templates alone cannot reliably extract the BAO scale.

\subsubsection{Completeness: $C$}
\label{sec:pure_sample}

Completeness, denoted by $C$, is the fraction of galaxies that FASHI successfully detects as a function of their integrated HI flux and line width, given the local detection sensitivity of the survey. 
That is, $0 \le  C \le 1$ quantifies the correction that must be applied to account for the sources missed by the detection pipeline~\cite{fashi_DR2}. 
All the analyses shown in previous sections considers the complete sample, where 
$C \ge 0$.

In this sub-section we assess the robustness of our results against survey incompleteness, we repeated the main angular correlation analysis using a cleaned subsample defined by stringent quality and completeness criteria. We perform a consistency test considering the sub-sample obtained with $C \ge 0.5$. 
This criterion filtered the original catalog providing a sample with $101,408$ HI sources, exhibiting a median redshift of $\bar{z} = 0.031$. While the fractional sky coverage remains unchanged at $f_{sky} = 0.40$, this reduction in source count results in an updated surface number density of $\Sigma = 20,279~\text{sr}^{-1}$. When computing the theoretical error bars for this complete sample, we maintain the linear galaxy bias and sky fraction fixed to isolate the direct impact of the completeness cuts on the clustering measurement. 

\begin{figure}
\centering
\includegraphics[scale=0.6]{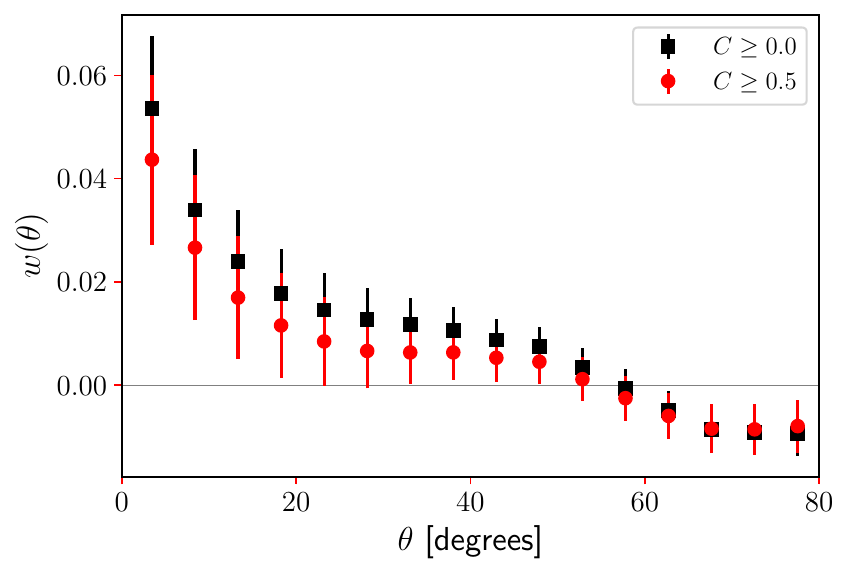}
\caption{The 2PACF $w(\theta)$ for the sample with $C\ge 0$ (black square points), compared against the sample with $C\ge0.5$ (red circle points).}
    \label{fig:pure_sample}
\end{figure}

The resulting 2PACF for this subsample are presented in Figure~\ref{fig:pure_sample}. Overall, the measurements remain remarkably consistent with those obtained from the baseline sample, demonstrating the strong stability and robustness of our clustering analysis against completeness and quality cuts. 

\subsubsection{Lognormal simulations}
\label{sec:glass_sims}

The statistical uncertainty in the 2PACF measurement is expressed by 
equation~\ref{eq:cov_matrix}. 
To robustly extract cosmological information from this signal, mock catalogs are essential. 
These simulations allow us to properly account for the survey's specific footprint, observational systematics, and the underlying gravitational evolution of density perturbations~\cite{Obuljen:2018kdy,Pfeifer:2023ntg}.

While lognormal simulations~\cite{Coles91,Agrawal17,Ramirez22} do not fully capture the complex, non-linear gravitational dynamics inherent to the local Universe, they serve as a practical first approximation for our study. We generate these mock catalogs using the lognormal approximation implemented in the \texttt{GLASS}\footnote{\url{https://glass.readthedocs.io}} framework \cite{Tessore23}. The lognormal model is a pretty efficient way to simulate how matter is distributed on large scales, and it completely avoids nonphysical negative densities. In this setup, you first generate a Gaussian random field from a chosen input power spectrum. Then you take the exponential of it to get a continuous lognormal density field, which gives a suitable approximation for mildly non-linear matter clustering. Finally, you draw discrete synthetic galaxies from that continuous field using Poisson sampling, specially tuned to match the radial selection function and angular mask of the FASHI survey.

The simulation setup defines the comoving radial resolution or redshift grid, $z_{\text{grid}}$, the choice of radial window function, the galaxy bias, $b$, the mask resolution, $n_{\text{side}}$, and the maximum multipole, $l_{\text{max}}$, employed to evaluate the angular power spectrum. To accurately reproduce the properties of our data, the code takes as input both the survey mask, handled using the \texttt{healpy} package\footnote{\url{https://healpy.readthedocs.io}}~\cite{2005ApJ...622..759G,Zonca2019}, and the observed distribution of galaxies per redshift interval and angular area ($dN/dz$). We adopt the same cosmological parameters as in our theoretical covariance matrix, using \texttt{CAMB}\footnote{\url{https://camb.readthedocs.io}}~\cite{Lewis:1999bs,Lewis:2026mif} with the Halofit model~\cite{Takahashi:2012em} to evaluate the nonlinear matter power spectrum. The mock galaxies are distributed across the redshift shells defined by $z_{\text{grid}} = [0.0 - 0.1, 0.015]$ using linear window functions and a constant bias of $b = 1.0$. In total, we generated $500$ mock catalogs. To compute the 2PACF for these realizations, we use the exact same random catalog constructed for the observed data. Table \ref{tab:sim_params} lists the set of input values used for generating the simulations.
\begin{table}[h]
    \centering
    \caption{Survey configuration parameters used to generate the set of $500$ mock catalogs.}
    \label{tab:sim_params}
    \begin{tabular}{c}
        \hline
        \hline
        \textbf{Survey configuration} \\ 
        \hline
        $z_{\text{grid}} = [0.0 - 0.1, 0.015]$ \\
        Linear window function \\
        $b = 1.0$ \\
        $n_{\text{side}} = 128$ \\
        $l_{\text{max}} = 256$ \\
        \hline
        \hline
    \end{tabular}
\end{table}

\begin{figure*}
    \centering
    \includegraphics[scale=0.6]{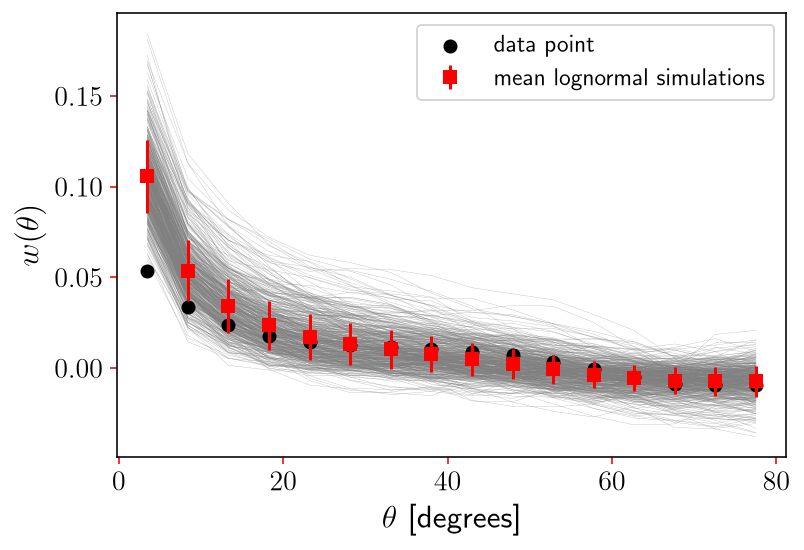}
    \includegraphics[scale=0.6]{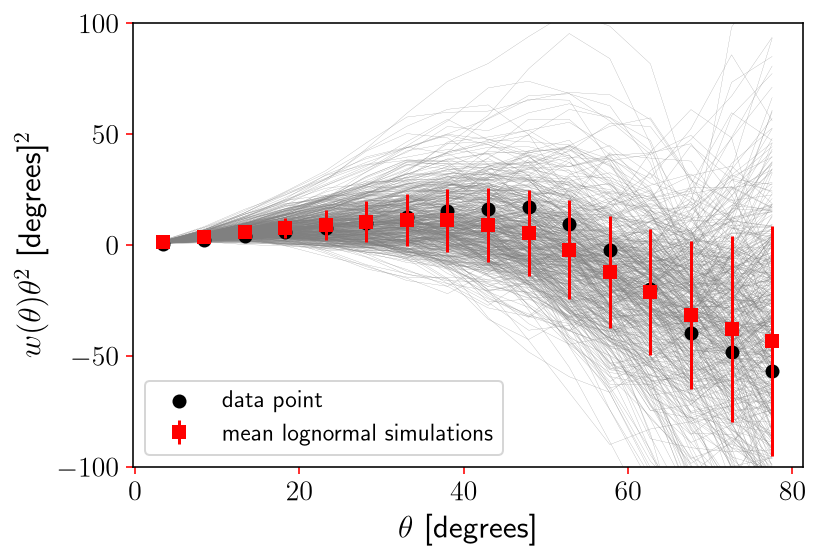}
    \caption{\textit{Left panel:} The 2PACF measured from the FASHI (black circles) compared to the ensemble of $500$ lognormal simulations. The gray lines indicate the individual mock realizations, while the red squares denote the mean of the mocks with error bars representing the $1\sigma$ standard deviation. \textit{Right panel:} The same data and simulations, but scaled by $\theta^2$ to highlight the structural features at large angular separations. While the lognormal model slightly over predicts the amplitude at small, it shows excellent agreement with the observational data at large scales.}
    \label{fig:glass_sim}
\end{figure*}

Figure~\ref{fig:glass_sim} presents the measured 2PACF from the FASHI data compared against the ensemble of $500$ lognormal mock catalogs. Note that the observational data points used for this comparison correspond to the main FASHI sample. The left panel displays the standard correlation function $w(\theta)$, while the right panel shows the signal scaled by $\theta^2$ to emphasize the large-scale clustering features. The individual mock realizations are shown as gray lines, and their mean is represented by the red squares, with error bars corresponding to the $1\sigma$ standard deviation of the ensemble. At small angular scales ($\theta \lesssim 15^\circ$), the simulations slightly over predict the clustering amplitude. On intermediate and large scales, the observational data points are in excellent agreement with the mock predictions, falling comfortably within the $1\sigma$ dispersion. This demonstrates that our lognormal simulations successfully capture the large scale clustering properties and the geometric footprint of the FASHI survey.

Regarding the broad bump observed in the correlation function, the mock catalogs provide an essential statistical baseline. As seen in Figure~\ref{fig:glass_sim}, the data points in this region fall entirely within the $1\sigma$ dispersion of the simulations. From a purely statistical standpoint, this indicates that the bump could be interpreted as a fluctuation driven by observational noise and cosmic variance. Nevertheless, the data exhibits a systematic trend: between $40^\circ$ and $60^\circ$, three consecutive data points lie above the mock mean, whereas between $60^\circ$ and $80^\circ$, three points fall systematically below it. Adopting a conservative stance, we conclude that while this coherent structure is present in the data, the current statistical uncertainties prevent us from unequivocally confirming it as a genuine cosmological signal over expected statistical fluctuations.

\subsubsection{Final Remarks}

In summary, we searched for the BAO signature studying the angular clustering of HI sources from the FASHI DR2 survey at the median redshift $\bar{z} = 0.034$, that is, in the extreme local Universe. By computing the angular 2-point correlation function, we observed hints of a broad acoustic feature peaking near $50^\circ$. An idealized thin-shell analysis would place the feature near $64^\circ$, while the thick-shell configuration that matches our sample (RSD $+$ $\Delta z = 0.09$) moves it to $\sim 52^\circ$ (see Figure~\ref{fig:proj_theory}). The observed maximum is therefore consistent with the thick-shell prediction and clearly displaced from the thin-shell value, indicating that line-of-sight projection is the dominant effect shaping the acoustic feature at these low redshifts.

Furthermore, we performed robustness tests varying the Zone of Avoidance footprint, 
as well as applying stringent completeness and quality cuts, and after these analyses 
we confirm the isotropic, stable, and cosmological nature of the signal. 
Comparisons with an ensemble of $500$ lognormal mock catalogs further validated 
our large scale measurements.

Because the data in analysis is located in a thick redshift shell, 
our approach theoretically investigates changes in BAO features resulting from 
modifications to the redshift shell thickness. 
According to this, our data analyses in a thick redshift shell confirms qualitatively 
that the BAO feature appears --indeed-- smeared, and quantitatively that a thick redshift shell 
produces an angular shift of the BAO signal of around $12^{\circ}$. 
A BAO detection with greater statistical significance would be possible with future FASHI data releases, due to improvements in both sky coverage and the number density of HI sources, reducing cosmic variance and shot noise, respectively. 
Overall, our study provides a strong observational baseline: they show that wide-field HI surveys like FASHI can trace primordial acoustic features even in the projection-dominated, low redshift regime. 
And, since BAO measurement at such low redshifts are very sensitive to $H_0$~\cite{Beutler11}, it could provide a standard ruler that offers an independent test to study the Hubble tension~\cite{DiValentino:2021izs,Cai:2026swf}.

\begin{acknowledgments}
FA acknowledges Fundação Carlos Chagas Filho de Amparo à Pesquisa do Estado do Rio de Janeiro (FAPERJ), Processo SEI-260003/001221/2025, for the financial support. CF acknowledges CAPES for the financial support. AB acknowledges a CNPq fellowship. 

This work has used the data from the Five-hundred-meter Aperture Spherical radio Telescope (FAST). FAST is a Chinese national mega-science facility, operated by the National Astronomical Observatories of Chinese Academy of Sciences (NAOC).
\end{acknowledgments}

\section*{Data Availability}
The data used in this work are available from the corresponding author upon reasonable request.

\bibliographystyle{apsrev4-1}
\bibliography{main}

\appendix

\section{Null Test and Shot Noise Limit}
\label{app:appendix_null_test}

To quantify the intrinsic noise floor and validate the stability of our estimator, we perform a null test using unclustered mock data. We generated a synthetic data catalog by drawing random positions within the exact spatial footprint of the FASHI DR2 subsample, thereby mimicking its angular geometry and number density but lacking any intrinsic spatial clustering. We then computed the 2PACF of this unclustered sample using the same random catalog and methodology applied to the actual data. To obtain robust statistics and map the distribution of the noise, this procedure was independently repeated $500$ times. 

\begin{figure}
    \centering
    \includegraphics[scale=0.55]{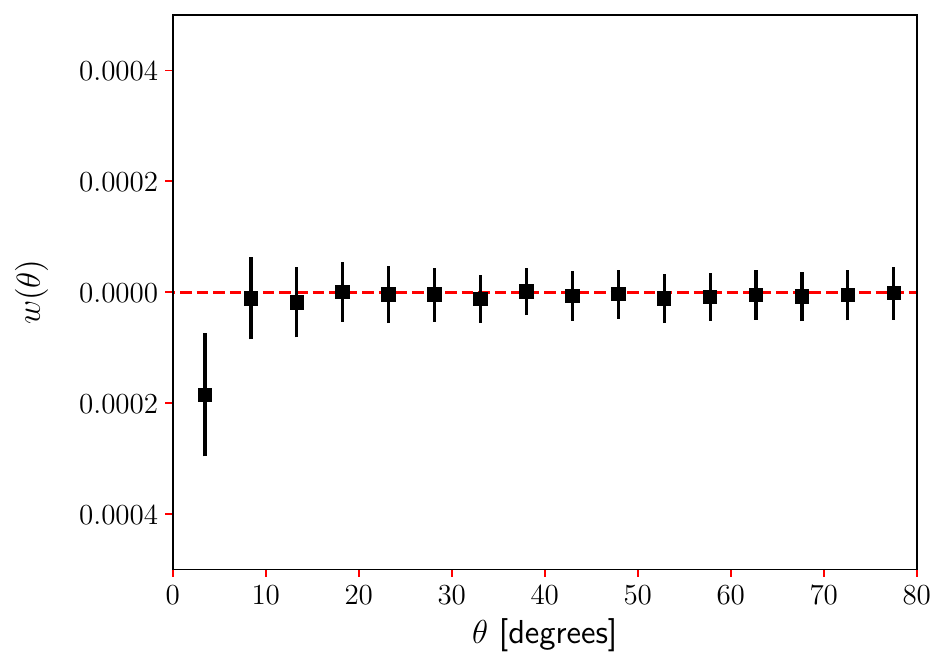}
    \caption{Null test of the 2PACF. The square markers represent the mean $w(\theta)$ computed from $500$ unclustered synthetic catalogs, mimicking the FASHI footprint and density. The error bars indicate the $1\sigma$ standard deviation of the 500 realizations. The red dashed line marks the theoretical expectation of zero correlation. The fluctuations remain on the order of $10^{-4}$, proving that the intrinsic shot noise limit of the estimator is negligible compared to the true physical signal.}
    \label{fig:shot_noise}
\end{figure}

As theoretically expected for an unclustered distribution, the resulting mean correlation is strictly consistent with zero across all angular scales. 
Figure~\ref{fig:shot_noise} presents the mean measurement and its $1\sigma$ dispersion. The amplitude of the shot noise fluctuations, as well as the statistical uncertainties, are bounded within an order of magnitude of $\sim 10^{-4}$. This noise floor is significantly smaller than the physical large scale clustering amplitude measured in the FASHI sample. This null test successfully demonstrates that our estimator is unbiased and that the observed large scale signal is not dominated by the catalog's shot noise.

\section{Surface pair density}
\label{app:density_pairs}

Since we are probing large angular scales, we must verify whether the pair count remains sufficiently high to ensure the measurements are statistically robust and not artifacts of the estimator due to boundary effects. Following \cite{Laurent_2016} and \cite{Ntelis_2017}, we use a simple and informative diagnostic to determine the maximum scale up to which the 2PACF can be reliably computed. We calculate the random pair density as a function of the angular scale. By normalizing this quantity at small separations, we can quantify the relative loss of accessible pairs as the scale increases. For our purely angular geometry, we define the scaled surface pair density, $\sigma(\theta)$, as 
\begin{equation}
\sigma(\theta) \equiv \frac{1}{\sin \theta}\,
\frac{dN}{d\theta} \,.
\end{equation}

\begin{figure}
    \centering
    \includegraphics[scale=0.6]{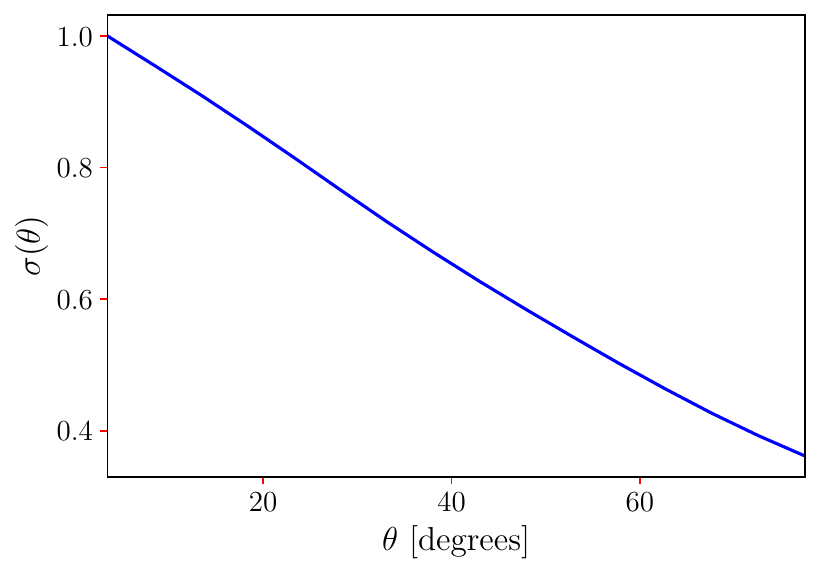}
    \caption{Normalized surface pair density of the random catalog, $\sigma(\theta)$, as a function of angular separation. The raw pair counts are scaled by the spherical geometric factor $(\sin \theta)^{-1}$ and normalized to unity at small scales. The gradual decline across the probed range confirms that boundary effects do not critically deplete the pair count, ensuring reliable clustering measurements up to $\theta = 80^\circ$.}
    \label{fig:density_pairs}
\end{figure}

Figure~\ref{fig:density_pairs} shows the resulting $\sigma(\theta)$ for our analysis. 
Along the angular range analyzed, $\theta \in [1^{\circ},80^{\circ}]$, we observe a maximum density drop of around 40\%. 
In contrast,~\cite{Laurent_2016} and~\cite{Ntelis_2017} established much lower operational thresholds in their respective works, truncating their analyses only when the scaled density dropped to 10\% and 1\%. 
Therefore, this test indicates that our results are highly robust and not significantly compromised by the loss of pairs at the largest probed scales.

\section{MCMC Posterior Distributions}
\label{sec:appendix_mcmc}

In this appendix we present the posterior distributions for the coefficients of the 5th-order polynomial used to perform the best-fit analyses of
the 2PACF (Section~\ref{sec:Results}), 
together with the coefficients corresponding to 
the quadratic baseline. 
The polynomial for this analysis is defined as
\begin{equation}
w(\theta) = \sum_{i=0}^{5} c_i \theta^i \,,
\end{equation}
where $\{ c_i, \,i=0,1,\cdots 5 \}$ represent the six free parameters of the fit,
and $\theta^i$ is the variable $\theta$ to the power $i$. 
The quadratic baseline has the same form, but is truncated at $i = 2$. 
We mapped the parameter space using the \texttt{emcee} package, assuming flat, uniform 
priors for all coefficients.

\begin{figure*}
\centering
\includegraphics[scale=0.45]{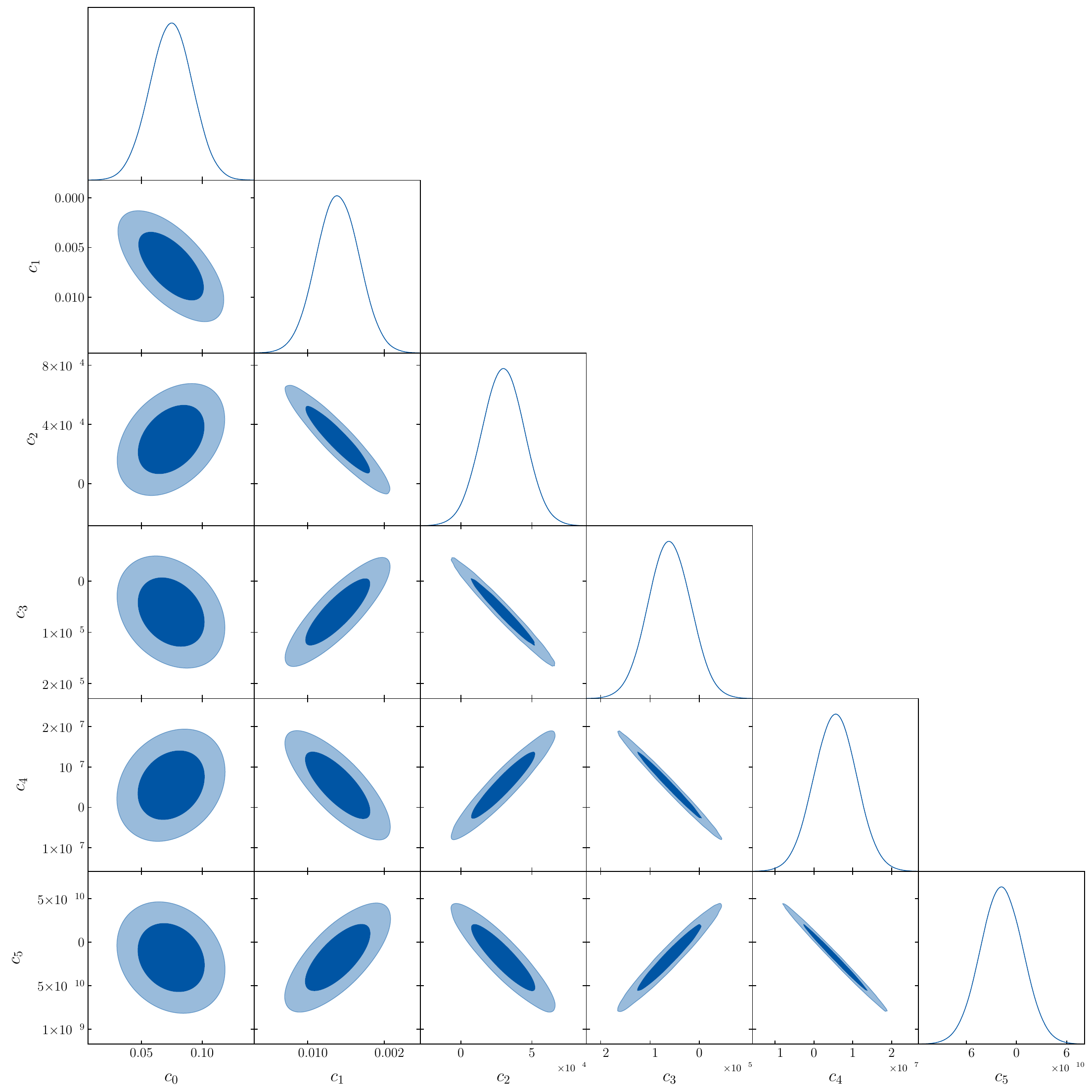}
\caption{
Marginalized 1D and 2D posterior distributions for the 5th-order polynomial coefficients ($c_0,c_1,c_2,c_3,c_4,c_5$) 
employed to best-fit the 2PACF, $w(\theta)$, obtained in the analysis of the FASHI data. 
The contour regions represent the $68\%$ and $95\%$ confidence levels. The elongated ellipses reflect the strong degeneracies among the higher-order coefficients, as expected for a polynomial fit to a smooth correlation function.}
\label{fig:polyfit5}
\end{figure*}

\begin{figure*}
\centering
\includegraphics[scale=0.45]{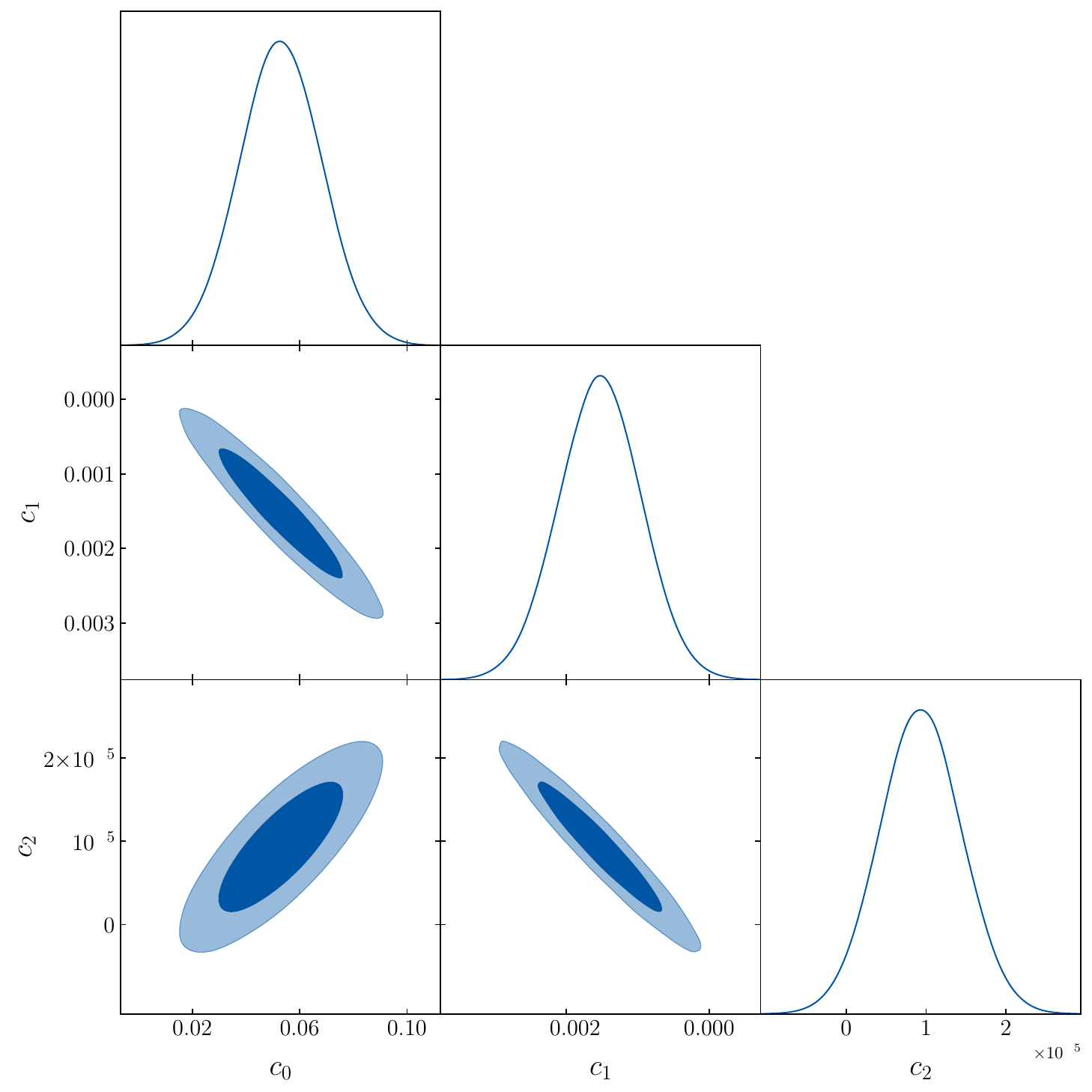}
\caption{
Marginalized 1D and 2D posterior distributions for the quadratic baseline coefficients ($c_0,c_1,c_2$) used to perform the best-fit of the 2PACF, $w(\theta)$, obtained in the analysis of the FASHI data. 
The contour regions represent the $68\%$ and $95\%$ confidence levels.}
\label{fig:polyfit2}
\end{figure*}

Figure~\ref{fig:polyfit5} displays the 1D marginalized posterior distributions and the 2D contour plots for the parameter covariances of the 5th-order polynomial fit, at the $68\%$ ($1\sigma$) and $95\%$ ($2\sigma$) confidence levels. The corresponding posteriors for the quadratic baseline are shown in Figure~\ref{fig:polyfit2}. 
The visualizations were generated using the \texttt{GetDist} Python package\footnote{\url{https://getdist.readthedocs.io}} \cite{Lewis:2019xzd}. 
The chains show no anomalous features and the projections onto individual parameters are well constrained, which allows for a reliable propagation of the uncertainties into the $dw/d\theta$ curve.

Table~\ref{tab:bao_significance} summarizes the best-fit parameters and the statistical comparison between the two models. The improvement in $\chi^2$ when moving from the quadratic to the 5th-order polynomial is $\Delta\chi^2 = 9.06$ for $\Delta\mathrm{dof} = 3$, corresponding to a $p$-value of $2.85 \times 10^{-2}$ and a nominal significance of
$1.9\,\sigma$. We note that the Bayesian Information Criterion gives
$\Delta\mathrm{BIC} = 0.74$, which is below the threshold of $2$ usually
required for positive evidence in favor of the more complex model. 
This reinforces the cautious interpretation of our findings, discussed in Section~\ref{sec:significance}: 
the polynomial comparison indicates the presence of non-quadratic structure in $w(\theta)$, 
but it does not, in itself, demonstrate a significant detection of a BAO signal.

\begin{table}[htbp]
\centering
\caption{Best-fit parameters for the 5th-order polynomial (Poly5) and the quadratic baseline (P2) models used to perform the best-fit of the 2PACF from FASHI data. 
We also display the statistical comparison quantifying the significance of non-quadratic structure in the measured 2PACF. 
The quoted uncertainties correspond to the $16$th and $84$th percentiles of the marginalized posteriors.}
\label{tab:bao_significance}
\begin{tabular}{lc}
\hline \hline
\textbf{Quantity} & \textbf{Value} \\
\hline
\multicolumn{2}{l}{\textbf{Poly5}} \\
$c_0$ & $(7.46 \pm 1.74) \times 10^{-2}$ \\
$c_1$ & $(-6.87 \pm 2.23) \times 10^{-3}$ \\
$c_2$ & $(2.98 \pm 1.49) \times 10^{-4}$ \\
$c_3$ & $(-6.08 \pm 4.29) \times 10^{-6}$ \\
$c_4$ & $(5.51 \pm 5.45) \times 10^{-8}$ \\
$c_5$ & $(-1.78 \pm 2.52) \times 10^{-10}$ \\
$\chi^2$ & $2.06$ \\
$\mathrm{dof}$ & $10$ \\
$\chi^2/\mathrm{dof}$ & $0.206$ \\
\hline
\multicolumn{2}{l}{\textbf{P2 (no-BAO)}} \\
$c_0$ & $(5.29 \pm 1.53) \times 10^{-2}$ \\
$c_1$ & $(-1.52 \pm 0.57) \times 10^{-3}$ \\
$c_2$ & $(9.31 \pm 5.12) \times 10^{-6}$ \\
$\chi^2$ & $11.12$ \\
$\mathrm{dof}$ & $13$ \\
$\chi^2/\mathrm{dof}$ & $0.855$ \\
\hline
\multicolumn{2}{l}{\textit{Model comparison} (\textbf{Poly5} \textit{vs.} \textbf{P2})} \\
$\Delta\chi^2$ & $9.06$ \\
$\Delta\mathrm{dof}$ & $3$ \\
$p$-value & $2.85 \times 10^{-2}$ \\
Stat. significance & $1.9\,\sigma$ \\
$\Delta\mathrm{AIC}$ & $3.06$ \\
$\Delta\mathrm{BIC}$ & $0.74$ \\
\hline \hline
\end{tabular}
\end{table}

\end{document}